\documentclass[nohyperref]{article}

\usepackage{microtype}
\usepackage{graphicx}
\usepackage{booktabs} 

\usepackage{hyperref}

\usepackage[accepted]{icml2023}

\usepackage{amsmath}
\usepackage{amssymb}
\usepackage{mathtools}
\usepackage{amsthm}

\usepackage[capitalize,noabbrev]{cleveref}

\theoremstyle{plain}

\theoremstyle{definition}

\theoremstyle{remark}

\usepackage[textsize=tiny]{todonotes}

\icmltitlerunning{EmoGen: Eliminating Subjective Bias in Emotional Music Generation}

\usepackage{xspace}
\usepackage[group-separator={,},group-minimum-digits={3}]{siunitx}
\usepackage{multirow} 
\usepackage{multicol}
\usepackage{bm}
\usepackage{makecell}
\usepackage{subcaption} 
\usepackage{stfloats}

\newcommand{\modelnamens}{\textsc{EmoGen}}
\newcommand{\modelname}{\modelnamens\xspace}

\makeatletter
\renewcommand*\subsectionautorefname{\S\@gobble}
\renewcommand*\sectionautorefname{\S\@gobble}
\makeatother

\begin{document}

\twocolumn[
\icmltitle{EmoGen: Eliminating
Subjective Bias in Emotional Music Generation}



\icmlsetsymbol{equal}{*}

\begin{icmlauthorlist}
\icmlauthor{Chenfei Kang}{equal,SJTU}
\icmlauthor{Peiling Lu}{equal,MS}
\icmlauthor{Botao Yu}{NJU}
\icmlauthor{Xu Tan}{MS}
\icmlauthor{Wei Ye}{Peking}
\icmlauthor{Shikun Zhang}{Peking}
\icmlauthor{Jiang Bian}{MS} \\
\vspace{0.2cm}
\url{https://github.com/microsoft/muzic}
\end{icmlauthorlist}

\icmlaffiliation{SJTU}{Shanghai Jiao Tong University, China}
\icmlaffiliation{NJU}{Nanjing University, China}
\icmlaffiliation{MS}{Microsoft Research Asia}
\icmlaffiliation{Peking}{National Engineering Research Center for Software Engineering, Peking University, China}

\icmlcorrespondingauthor{Xu Tan}{xuta@microsoft.com}

\icmlkeywords{Music generation, music emotion, music attributes, supervised clustering, self-supervised learning}

\vskip 0.3in
]



\printAffiliationsAndNotice{\icmlEqualContribution} 

\begin{abstract}
Music is used to convey emotions, and thus generating emotional music is important in automatic music generation. 
Previous work on emotional music generation directly uses annotated emotion labels as control signals, which suffers from subjective bias: different people may annotate different emotions on the same music, and one person may feel different emotions under different situations.
Therefore, directly mapping emotion labels to music sequences in an end-to-end way would confuse the learning process and hinder the model from generating music with general emotions.
In this paper, we propose \modelname, an emotional music generation system that leverages a set of emotion-related music attributes as the bridge between emotion and music, and divides the generation into two stages: emotion-to-attribute mapping with supervised clustering, and attribute-to-music generation with self-supervised learning.
Both stages are beneficial: 
in the first stage, the attribute values around the clustering center represent the general emotions of these samples, which help eliminate the impacts of the subjective bias of emotion labels; in the second stage, the generation is completely disentangled from emotion labels and thus free from the subjective bias.
Both subjective and objective evaluations show that \modelname outperforms previous methods on emotion control accuracy and music quality respectively, which demonstrate our superiority in generating emotional music. Music samples generated by \modelname are available via this link$\footnote{\label{demo}\url{https://ai-muzic.github.io/emogen/}}$, and the code is available at this link$\footnote{\label{code}\url{https://github.com/microsoft/muzic/}}$.

\end{abstract}

\section{Introduction}  \label{sec:intro}

With the development of deep learning, automatic music generation is developing rapidly and attracting more and more interest \cite{hernandez2022music,shih2022theme,yu2022museformer}. 
Due to the importance of emotions for music, emotional music generation is an important and practical task, yet it is still under-explored.

Previous work, according to the way of applying emotion signals, can be divided into two types. 
The first type is to convert emotion labels as embeddings and take them as model input \cite{madhok2018sentimozart, hung2021emopia, pangestu2021generating, sulun2022symbolic, grekow2021monophonic}. 
The second type is to train an emotion classifier and apply it at either model output to guide the decoding process \cite{ferreira2021learning, ferreira2020computer, ferreira2022controlling, bao2022generating}, or latent space of variational autoencoders \cite{tan2020music} and generative adversarial networks \cite{tseng2021extending} to constrain the distribution of latent vectors. 

However, both the above two types of work directly use emotion labels as the control signals to generate music sequences in an end-to-end way, which is suboptimal. Emotion labels given by data annotators can be influenced by both objective and subjective factors. Objective factors like tempo and note density are highly associated with music emotions. As for subjective factors, the perceived emotions are highly related to social identities, personalities, instant emotional states of listeners, etc. For example, it is highly possible that a listener thinks a happy song is sad when he/she is in an upset state. 
Due to this subjectivity of human emotions, different data annotators may give different emotion labels to the samples with the same emotion, which results in subjective bias in emotion labels.
With the inconsistent emotion labels, it is hard for those end-to-end methods to learn the relationship between emotion and music sequences, and accordingly, the models can be deficient in generating music that exactly matches the desired emotion.

In this paper, we propose \modelname, an emotional music generation system that can eliminate the impacts of subjective bias of emotion labels. Instead of directly mapping emotion labels to music sequences in an end-to-end way, we leverage a set of music attributes that are highly correlated with emotions as a bridge and break down this task into two stages: emotion-to-attribute mapping with supervised clustering, and attribute-to-music generation with self-supervised learning.

Specifically, to bridge the gap between emotions and music, the attributes need to be highly correlated with emotions. We design the attribute set by training an emotion classifier on a labeled dataset and selecting those attributes whose feature importance is high.

In the emotion-to-attribute mapping stage, we map the emotion to the attribute values of a sample closest to the clustering center, which is obtained by clustering samples with emotion labels and calculating the average attribute values from each cluster. This clustering process is supervised since we use emotion labels to cluster samples into emotion categories. By this supervised clustering, mapped attribute values can represent the general emotion from the samples around the clustering center. Thus, the problem of subjective bias from emotion labels can be eliminated. 
 
In the attribute-to-music generation stage, we extract the attribute values from music sequences and train a Transformer-based model with these attributes as the control signals in a self-supervised way. The values of the attributes can be directly extracted from music sequences, so this generative model can learn the relationship between control signals and music without requiring any labeled data. Since this process is completely disentangled from emotion labels, it is not influenced by the subjective bias of emotion labels. With the benefits of the two stages based on supervised clustering and self-supervised learning on avoiding the subjective bias, \modelname can achieve a more precise emotion control in emotional music generation. 

The main contributions of this work are as follows: 
\begin{itemize}
\item We propose \modelname, an emotional music generation system that can eliminate subjective bias from emotion labels, which leverages emotion-related attributes as a bridge to generate music with desired emotions by two stages: emotion-to-attribute mapping with supervised clustering and attribute-to-music generation with self-supervised learning.
\item Experimental results show that \modelname outperforms previous methods on emotion control accuracy and music quality. Experiments also demonstrate the ability of \modelname to eliminate subjective bias in emotion labels.

\end{itemize}

\section{Related Work}

\subsection{Emotional Music Generation}
Emotion-conditioned music generation is developing rapidly in the age of deep learning. According to the way of applying emotion signals, previous work can be divided into two types. The first type is to convert emotion labels as embeddings and take them as model input
\cite{madhok2018sentimozart, hung2021emopia, pangestu2021generating, sulun2022symbolic, grekow2021monophonic}. 
\citeauthor{madhok2018sentimozart} generate emotional music based on the one-hot emotion label. Some work \cite{hung2021emopia, pangestu2021generating} add extra emotion tokens into MIDI events to generate music with specific emotions. \citeauthor{sulun2022symbolic} control music generation conditioned on continuous-valued valence and arousal labels.
The second type is to train an emotion classifier and apply it at either model output through heuristic search methods guiding the decoding process \cite{ferreira2021learning, ferreira2020computer, ferreira2022controlling, bao2022generating}, or latent space of variational autoencoders \cite{tan2020music} or generative adversarial networks \cite{tseng2021extending} to constrain the distribution of latent vectors. \citeauthor{ferreira2021learning} use Genetic Algorithm to optimize the weights of the Long Short-Term Memory (LSTM) to generate music with desired emotions. Some work \cite{bao2022generating, ferreira2020computer, ferreira2022controlling} apply search algorithm (e.g., beam search and tree search) to direct music generation with desired emotions. However, both of the above two types directly use emotion labels as the control signals to guide music generation, which ignores the impact of subjective bias of emotion labels as discussed in \autoref{sec:intro}. Therefore, it is difficult for existing methods to generate music that matches the desired emotion.

\subsection{Attribute-Based Controllable Music Generation}

Music attributes are extracted from music sequences and can be manipulated to control music generation. Previous work attempt to leverage these attributes for controlling the music generation process. These works \cite{tan2020music, kawai2020attributes, zhao2022domain} extract attributes like rhythm density, pitch and rhythm variability, and chords and apply a VAE-based framework to control music generation by music attributes. A discriminator is used to control the hidden space distribution to satisfy the attribute conditions. \citeauthor{wu2021musemorphose} propose MuseMorphose, which adds rhythmic and polyphony intensity into the latent space of VAE to control music generation. \citeauthor{von2022figaro} propose FIGARO, which designs expert description (such as note density, mean pitch, etc.) and learned description (latent representation) to control music generation with a VQ-VAE system. Directly using these attributes for emotional music generation is not enough, since they either do not consider building relationships between emotions and attributes, or fail to construct a concrete correlation between attributes and emotions, which may result in poor controlling accuracy.

\section{Method} \label{sec:method}

\begin{figure}[t!]
\begin{subfigure}[b]{\columnwidth}
    \centering
    \includegraphics[width=\textwidth]{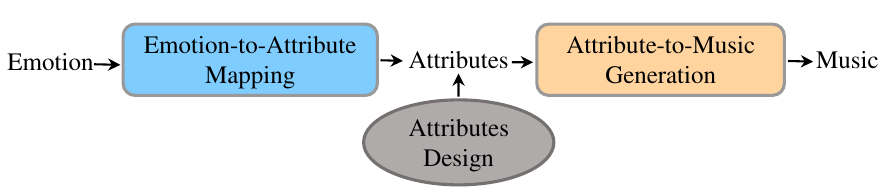}
    \caption{Pipeline of \modelname.}
    \label{fig:pipline}
\end{subfigure}
\begin{subfigure}[b]{\columnwidth}
    \includegraphics[width=\textwidth]{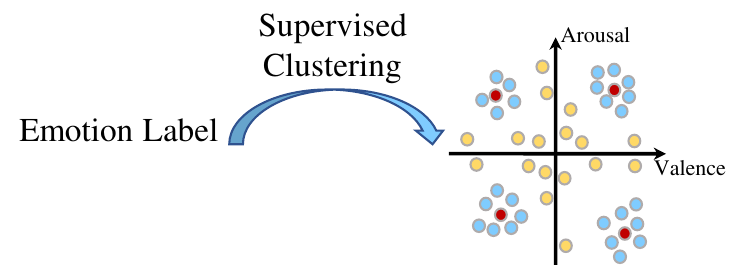}
    \caption{Emotion-to-attribute mapping with supervised clustering. The emotion space is divided by arousal and valence into four quadrants based on Russell's 4Q model \cite{russell1980circumplex}. Mapped attribute values from emotion labels that represent general emotions are marked with red dots, while attribute values from emotion labels that contain subjective bias are marked with yellow dots.}
    \label{fig:stage1}
\end{subfigure}
\begin{subfigure}[b]{\columnwidth}
    \includegraphics[width=\textwidth]{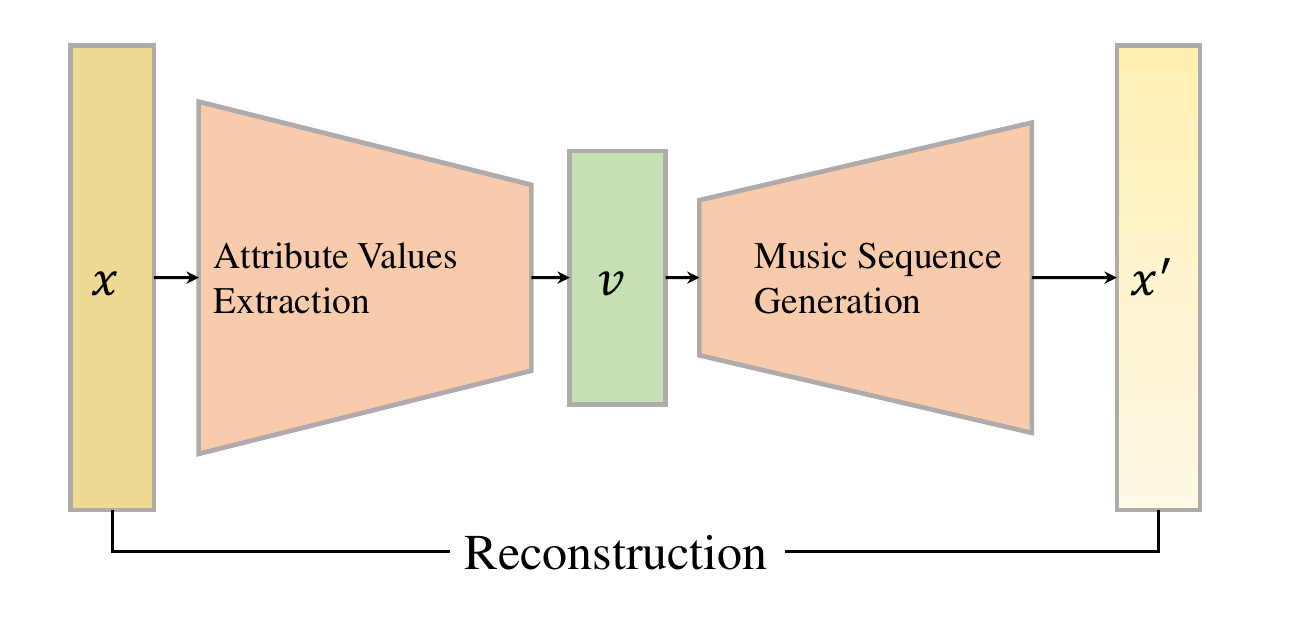}
    \caption{Training process of attribute-to-music generation in self-supervised learning. $x$ represents the target music sequences, $x'$ represents the generated music sequences, and $v$ represents the $d$-dimension vector of extracted attribute values from target music sequences.}
    \label{fig:stage2}
\end{subfigure}
\caption{Overview of \modelname.}
\end{figure}

\autoref{fig:pipline} shows the pipeline of \modelname,  which contains two stages: emotion-to-attribute mapping, and attribute-to-music generation, with a set of designed attributes as a bridge.
In attribute design, we enumerate and select the set of attributes that are highly correlated with emotions, which can help build a consistent relationship between emotions and music.
For the emotion-to-attribute mapping stage, we get the mapped attribute values by choosing the values closest to the clustering center. The mapped attribute values can well represent the general emotions, so as to alleviate the subjective bias from emotion labels.

For the attribute-to-music generation stage, the attribute values directly extracted from music sequences are used as control signals for training an autoregressive Transformer based model to generate corresponding music sequences in a self-supervised way. By disentangling the generation process from emotion labels, we can avoid the subjective bias from emotion labels to achieve better control in emotional music generation. We discuss the merits of our system in \autoref{sec:merits}.

\subsection{Emotion-Related Attribute Design} \label{sec:attr_selection}

Instead of generating music sequences with the conditions of emotion labels, where subjective bias exists, we introduce emotion-related attributes to bridge the gap between emotions and the corresponding music. Compared with emotion labels, these objective attributes tell exactly what the corresponding music should be. For example, the tempo value tells what the duration of a beat is, while the key scale states what a set of notes can be used. By directly extracting the values of these attributes from music sequences, we can help build an explicit relationship between emotions and music.
Specifically, we collected music attributes from low-level features like pitch statics, chords and vertical intervals, rhythm, and dynamics to high-level features like melodic intervals, instrumentation, and musical texture \cite{mckay2018jsymbolic}.

However, since many of them are irrelevant to emotions, directly using all of them would introduce a lot of noise. Thus, we select attributes that are highly correlated with emotions by training a Random Forest (RF) \cite{ho1995random} classifier on an emotion-annotated dataset, then picking up the top-$k$ attributes according to the ranking of feature importance as the final attributes set. Through this process, the designed attributes can represent emotional information and help control the music generation. Please refer to \autoref{app:attributes} for details of these designed attributes.

\subsection{Emotion-to-Attribute Mapping} \label{sec:stage1}

To generate music with the desired emotion based on the emotion-related attributes, the emotion label is mapped to the attribute values that represent the general emotion with supervised clustering as shown in \autoref{fig:stage1}.

Specifically, we first extract the values of the selected attributes for each sample in an emotion-annotated dataset. 
Based on the emotion labels given by the dataset, among the samples of each emotion label, we calculate the mean value for each attribute to obtain the center. 
Then, the attribute values of the sample that is the closest to the center are used to represent the features of the emotion.

This process is supervised since emotion labels are used as clustering guidance to group samples into categories. This is also a clustering process since the samples are grouped in such a way that samples in the same group share similar emotional information, while samples in different groups convey distinct emotions.
Through this supervised clustering method, the obtained attribute values should be able to represent the general emotion given this emotion label and avoid the subjective bias coming from emotion labels.

\subsection{Attribute-to-Music Generation} \label{sec:stage2}

Attribute values can be easily extracted from music sequences, which is much more precise for controlling music generation. The training process is shown in \autoref{fig:stage2}, we extract the values of the emotion-related attributes from the target music sequence and represent them in a $d$-dimension vector, we then take this vector as control signals into an autoregressive Transformer based model for generating the corresponding music sequence. The model is trained with the mapped attribute vectors as supervisory signals in a self-supervised way. Through this self-supervised learning step, the learned Transformer model is able to generate music whose attributes are controlled by the input attribute values. When inference, the attribute values mapped in the emotion-to-attribute mapping stage are leveraged as the control signals to guide the music generation process. Since the generation process is completely disentangled from emotion labels, it is not affected by the subjective bias from emotion labels.

\subsection{Merits of \modelname} \label{sec:merits}

This proposed framework is beneficial for generating music with desired emotion in the following aspects:

\begin{itemize}
    \item \textit{Ability to eliminate subjective bias.} By leveraging the supervised clustering and self-supervised learning paradigm in the two stages, \modelname can eliminate subjective bias from emotion labels to achieve better emotion control accuracy. Specifically, by mapping emotions to emotion-related attributes with supervised clustering, we obtain the values of attributes on behalf of the general emotion. By training the autoregressive Transformer based model with attribute values as control signals in a self-supervised way, we disentangle emotion labels from the generation process and build an explicit relationship between control signals and music sequences. The emotion labels are not directly used in the whole generation process so that we can avoid the subjective bias that exists in emotion labels.
    \item \textit{Ability to precisely control generation process.} Music attributes are good tools to concretely direct the generation. A single emotion label is too ambiguous to define what the corresponding music should be. For example, it is hard to define what a piece of happy music should be like. In contrast, music attributes are concrete to designate specific aspects of music \cite{tan2020music,wu2021musemorphose,von2022figaro,di2021video,chen2020music}. For example, the tempo value tells exactly the duration of one beat, and the type of scale determines what sets of notes are used in generated music. By simply manipulating the values of the music attributes, we can precisely control the generated music.
    \item \textit{Ability to be free from labeled data.} \modelname can generate emotion-conditioned music without requiring any emotion annotation. Manual annotation is expensive, and there are only a few datasets \cite{hung2021emopia} that contain emotion annotations. Unlike the previous methods that require emotion-music paired data for training the generative model, in \modelname, emotion labels are only used to determine the emotion-related attributes and the mapped attribute values that represent the general emotion. Once they are determined, they will not be changed. After that, we can simply extract the attribute values on an arbitrary dataset to train the generative model on it with self-supervised learning. Therefore, \modelname can be used to generate emotion-conditioned music even if the dataset has no emotion annotations.
\end{itemize}

\section{Experiment} \label{sec:exp}
In this section, we first introduce the experiment setup (\autoref{sec:exp_setup}), followed by the comparison with previous methods. After that, we give a comprehensive discussion on how \modelname eliminates the subjective bias of emotion labels. Then we show the comprehensive analysis of \modelname. Finally, we show the results of applying the framework of \modelname to other arbitrary datasets with no annotations.

\subsection{Experiment Setup}
\label{sec:exp_setup}

\paragraph{Datasets} We use altogether three datasets including one emotion-labeled dataset namely EMOPIA \cite{hung2021emopia}, and two unlabeled datasets namely Pop1k7 \cite{hsiao2021compound} and LMD-Piano, where LMD-Piano is constructed by using the samples that only contain piano tracks from the Lakh MIDI (LMD) dataset \cite{raffel2016learning}. The information of these datasets is shown in \autoref{tab:datasets}. EMOPIA uses Russell's 4Q model \cite{russell1980circumplex} as the emotion classification criterion, which is also leveraged in our evaluation process. EMOPIA with emotion labels is used to determine the designed attributes and the mapped attribute values in the emotion-to-attribute stage. Once they are determined, they are kept unchanged and emotion labels will
not be used. It is also used in the fine-tuning stage when compared with previous methods. Pop1k7 and LMD-Piano are used for the pre-training stage when compared with previous methods.
We randomly split each dataset by $8/1/1$ for training/validation/test, respectively.
\begin{table}[t]
    \centering
    \caption{\centering The information of training datasets.}
    \begin{tabular}{c|ccc}
        \toprule
        Name & Music type  & Size & Label type \\
        \midrule
        EMOPIA & Piano & \num{1078} & Russell's 4Q \\
        Pop1k7  & Piano & \num{1748} & None \\
        LMD-Piano & Piano & \num{22643} & None \\
        \bottomrule
    \end{tabular} 
    \label{tab:datasets}
\end{table}

 \paragraph{System Configurations} 
 We use a REMI-like \cite{huang2020pop} representation method to convert MIDI into token sequences.
 We apply jSymbolic \cite{mckay2018jsymbolic} to extract attribute values from music, and train the Random Forest classifier on EMOPIA, then select the top-$100$ attributes that are most related to emotions as described in \autoref{sec:attr_selection}.
 In the emotion-to-attribute mapping stage (\autoref{sec:stage1}), supervised clustering is implemented on EMOPIA. In the attribute-to-music generation stage (\autoref{sec:stage2}), we binarize the mapped attribute vector with the median, which is then fed into a $2$-layer feed-forward network to obtain the attribute embedding. It is then added onto the token embedding at the input of an autoregressive Transformer model.
 We leverage Linear Transformer \cite{katharopoulos2020transformers} as the backbone model, which consists of \num{6} Transformer layers with causal attention and \num{8} attention heads. The attention hidden size is \num{512} and FFN hidden size is \num{2048}. The max sequence length of each sample is \num{1280}. During training, the batch size is set to \num{8}. We use Adam optimizer \cite{kingma2014adam} with $\beta_1 = 0.9$, $\beta_2=0.98$ and $\epsilon=10^{-9}$. The learning rate is $1 \times 10^{-4}$ with warm-up step $16000$ and an inverse-square-root decay. The dropout rate is $0.1$. During inference, we apply top-$p$ sampling with ratio $p=0.9$ and temperature $\tau=1.0$. 

\paragraph{Compared Methods} We compare \modelname with two representative methods of different emotion control manners. The first one is Conditional Sampling (CS) \cite{hung2021emopia}, which uses extra emotion tokens in the model input as the emotion conditions.  The other one is Predictor Upper Confidence for Trees (PUCT) \cite{ferreira2022controlling}, which uses an emotion classifier and a music discriminator trained on labeled data to direct the inference process.

 \paragraph{Evaluations and Metrics}

We conduct both subjective and objective evaluations to evaluate \modelname. Each model is applied to generate \num{1000} music pieces, with \num{250} for each of the four emotions.
In subjective evaluation, human scorers are asked to rate each music piece.
We report the following subjective metrics: 1) Subjective accuracy: Whether the perceived emotion by subjects is consistent with the emotion label. It represents emotion controllability. 2) Humanness: How similar it sounds to the music composed by a human. It represents the music quality. 3) Overall: an overall score.
In objective evaluation, following \cite{ferreira2022controlling}, we use a Linear Transformer-based emotion classifier trained on EMOPIA to predict the emotion label of each generated music piece. Then we calculate objective accuracy by comparing the emotion input for generating music with the predicted emotion class by this classifier. We report the objective accuracy of the classification, which functions as a supplementary metric to the subjective accuracy.
For more details about the human rating process and the evaluation metrics, please refer to Appendix \autoref{app:main_results}.

\subsection{Comparison with Previous Methods} \label{sec:main_results}

In order to conduct thorough evaluations, we design two training settings: 

1) Setting 1 (\texttt{S1}): To ensure the music quality of generated music, following previous work \cite{hung2021emopia, ferreira2022controlling, neves2022generating}, we pre-train the models on Pop1k7+LMD-Piano before fine-tuning on EMOPIA. 
For \modelname, we first pre-train the language model with designed attributes as control signals, then fine-tune it on EMOPIA with attributes as control signals. 
For CS, following the work of \cite{hung2021emopia}, we pre-train the language model with the control of the emotion token setting to ``\texttt{<None>}'' as a placeholder, followed by attributes. After pre-training, we finetune the model on EMOPIA with emotion tokens assigned to the placeholder, followed by attributes as control signals. 
For PUCT, we first pre-train the language model, then fine-tune it on EMOPIA with an extra classification head to get the emotion classifier. We train the music discriminator by fine-tuning the language model with an extra classification head to classify real/fake samples.
All of the methods are pre-trained on Pop1k7 and LMD-Piano.

2) Setting 2 (\texttt{S2}): The training methods in the above setting have their limitations in that they can only generate music similar to the dataset used in the fine-tuning stage. This constrains the ability of \modelname, which can naturally leverage arbitrary datasets for emotional music generation. To test this ability, we train the generative model on Pop1k7+LMD-Piano+EMOPIA in the attribute-to-music generation stage, and use the designed and mapped attributes for generating corresponding music with given emotions. Please note that since CS and PUCT require only labeled data in training, they cannot work in this setting. Therefore, we compare \modelname with the ground truth, the EMOPIA dataset.

\begin{table*}[t]
    \centering
     \caption{ The results of the subjective and objective evaluation. For humanness and overall scores, we report mean opinion scores with $95$\% confidence interval.}
    \begin{tabular}{cc|ccc|c}
    \toprule
                     &                    & \multicolumn{3}{c|}{Subjective}   & Objective  \\
   Setting    & Method                    & Accuracy$\uparrow$ & Humanness$\uparrow$ & Overall $\uparrow$ & Accuracy $\uparrow$  \\ \midrule
     EMOPIA & Ground truth & $0.433$ & $4.26$ $\pm$ $0.15$ & $4.19$ $\pm$ $0.15$ &  $0.628$  \\ \midrule
\multirow{3}{*}{
\thead{\texttt{S1}: Pre-training on Pop1k7+LMD-Piano, \\fine-tuning on EMOPIA}} & CS  & $0.250$    &   $3.48 \pm 0.22$      &  $3.59\pm0.22$  &   $0.439$    \\
                             & PUCT     &  $0.150$    &  $3.24 \pm 0.24$    &  $3.26\pm0.23$ &  $0.260$  \\
                             & \modelname &   \bm{$0.650$}  &   \bm{$3.54$} $\pm$ \bm{$0.19$}   & \bm{$3.59$} $\pm$ \bm{$0.18$}  &  \bm{$0.715$}   \\ \midrule

     \thead{\texttt{S2}: Training on Pop1k7+LMD-Piano+EMOPIA} & \modelname &  $0.550$      &  $3.67$$\pm$ $0.20$  & $3.65$ $\pm$ $0.20$    &   $0.658$    \\ 
    \bottomrule
    \end{tabular}
        \centering
        \label{tab:main_results}
\end{table*}

 The results are shown in \autoref{tab:main_results}. We can observe that: 
 1) Compared with CS and PUCT, \modelname achieves better performance on all the metrics in \texttt{S1}. Particularly, \modelname has much better emotion controllability on both the subjective and objective accuracy. It demonstrates the superiority of \modelname in generating music with designated emotion. Besides, the higher humanness and overall score of \modelname indicate that \modelname is capable of improving the music quality.
 
 2) In \texttt{S2}, \modelname achieves higher performance on all the metrics than CS and PUCT in \texttt{S1}. It demonstrates that \modelname can not only leverage an arbitrary dataset for emotional music generation but also have fairly good controllability and humanness. We will also show its ability on a dataset of more diverse and multi-instrument music in \autoref{sec:ext_topmagd}.
 
 3) The accuracy of \modelname is even higher than that of the ground truth (EMOPIA). This shows that, on the one hand, there are samples with ambiguous emotion labels in the labeled dataset that affect the judgment of their emotions. On the other hand, the two-stage framework of \modelname, especially the mapped attribute values from emotions determined by supervised clustering in the emotion-to-attribute mapping stage, can help avoid the subjective bias from emotion labels.
 Thus, we choose \texttt{S2} as our basic framework and use it to do further analysis of \modelname.

\subsection{Verification on Eliminating Subjective Bias} \label{sec:bias}
To demonstrate that \modelname can eliminate subjective bias in emotion labels, we conduct experiments to show 1) the existence of subjective bias in a labeled dataset and 2) \modelname's ability to eliminate subjective bias. We use the subjective and objective accuracy described above as the metrics.

\begin{figure}[ht]
\begin{subfigure}[b]{0.49\columnwidth}
    \centering
    \includegraphics[width=\textwidth]{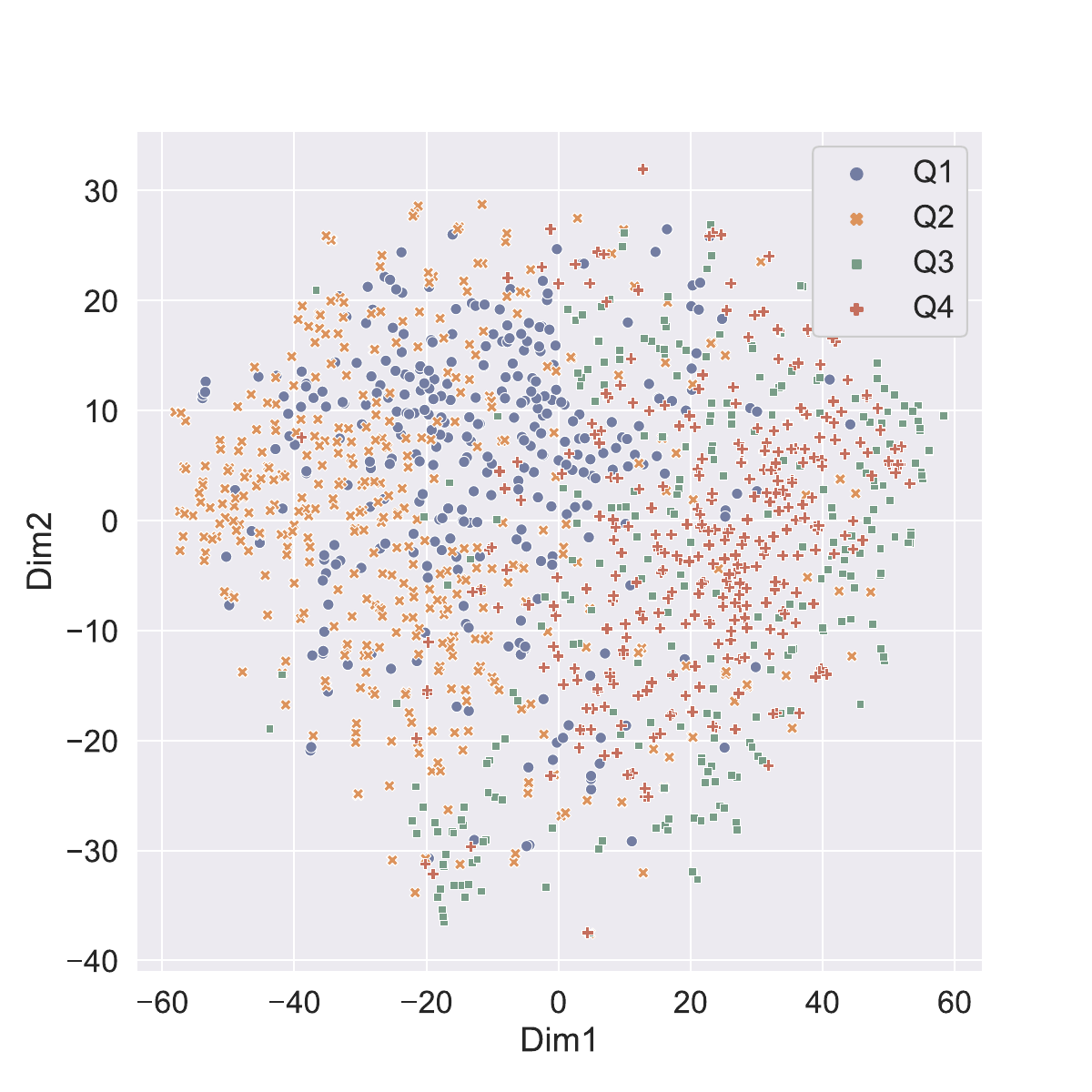}
    \caption{EMOPIA}
    \label{fig:t_sne_emopia}
\end{subfigure}
\begin{subfigure}[b]{0.49\columnwidth}
    \centering
    \includegraphics[width=\textwidth]{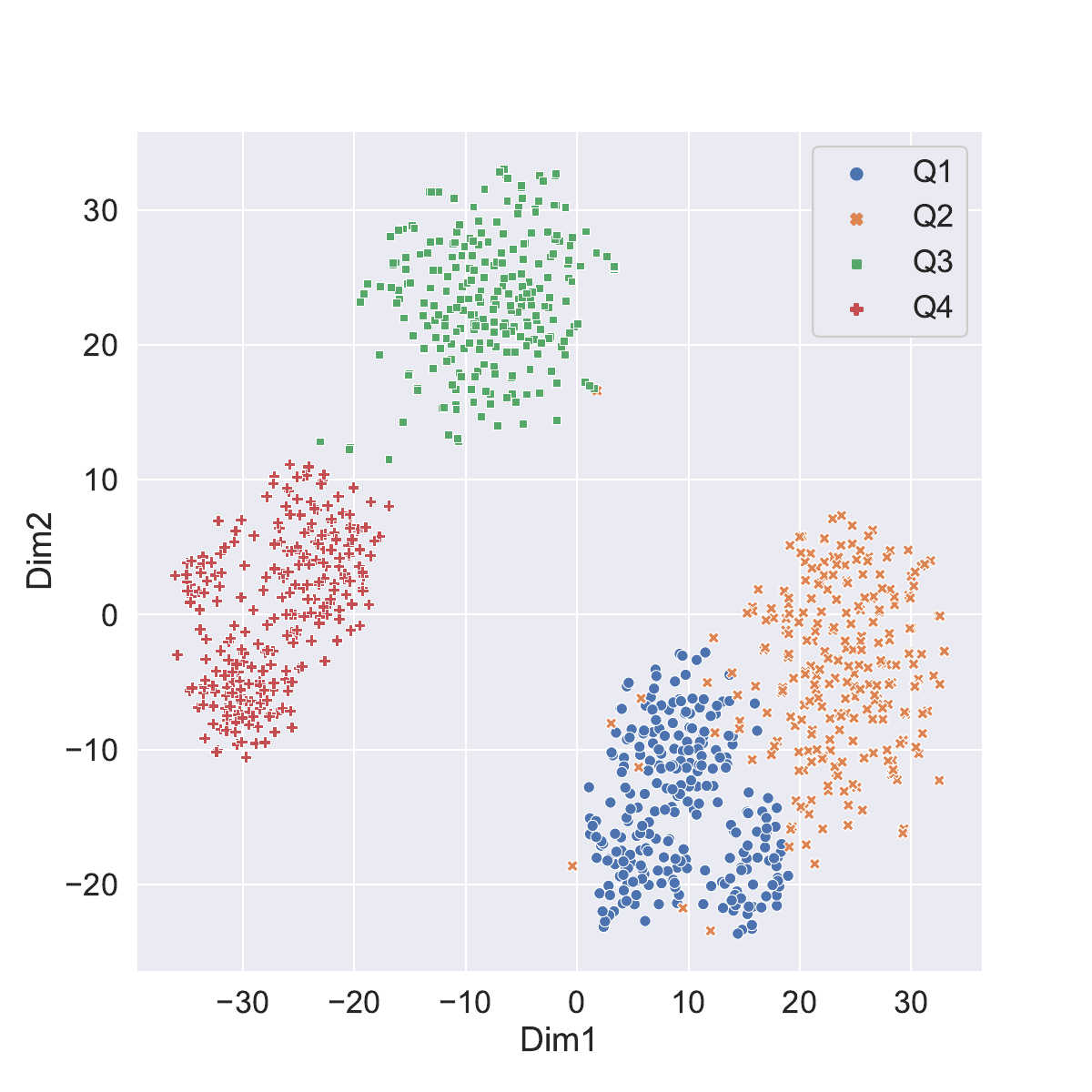}
    \caption{\modelname}
    \label{fig:t_sne_emogen}
\end{subfigure}

\caption{T-SNE visualization of attribute vectors from samples in EMOPIA and samples generated by \modelname. "$Qi$" represents the $i$-th quadrant of Russell's 4Q model.}
\label{fig:t_sne}
\end{figure}

\begin{figure}[ht]
\begin{subfigure}[b]{0.49\columnwidth}
    \centering
    \includegraphics[width=\textwidth]{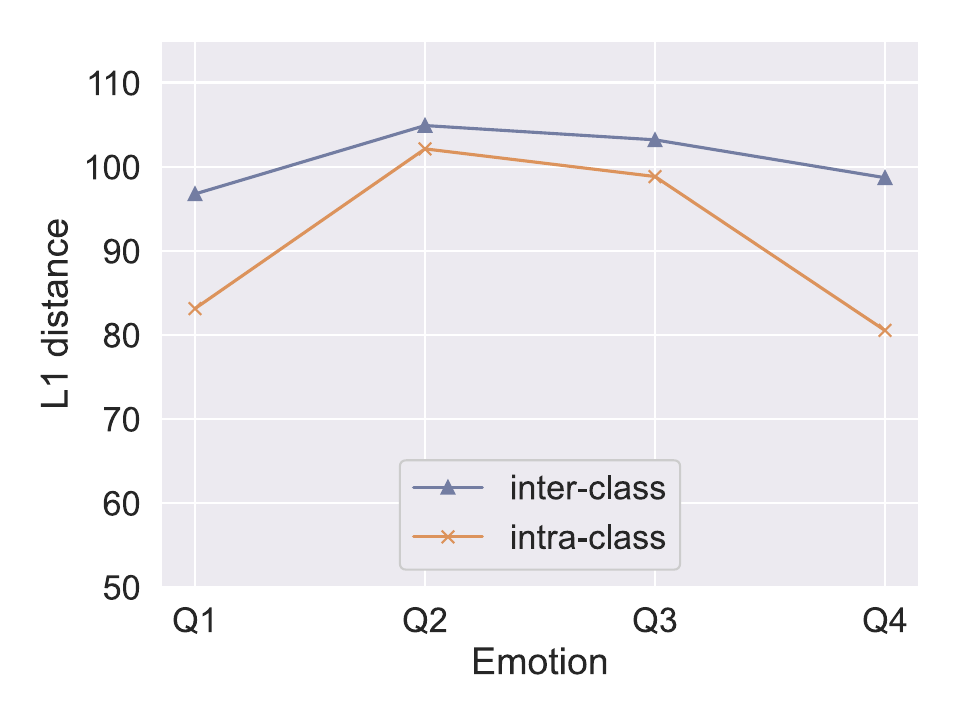}
    \caption{EMOPIA}
    \label{fig:l1_distance_emopia}
\end{subfigure}
\begin{subfigure}[b]{0.49\columnwidth}
    \centering
    \includegraphics[width=\textwidth]{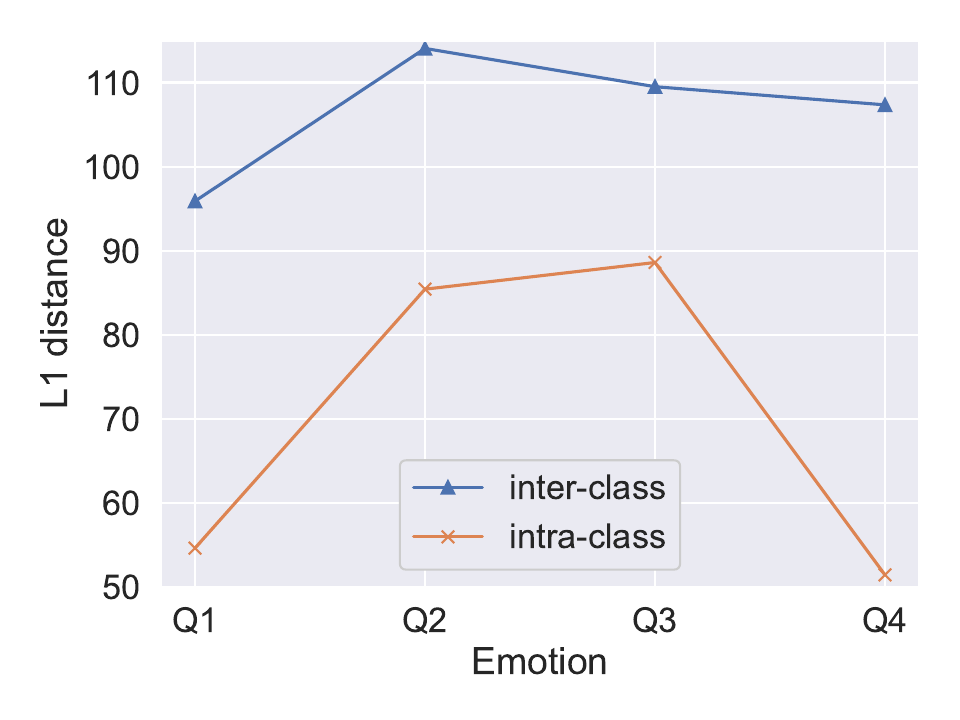}
    \caption{\modelname}
    \label{fig:l1_distance_emogen}
\end{subfigure}

\caption{Intra and inter L1 distance of attribute vectors from samples in EMOPIA and samples generated by \modelname. "Intra-class" means the average distance of attribute vectors with the same emotional labels and "inter-class" means that with different emotional labels.}
\label{fig:l1_distance}
\end{figure}

\paragraph{Existence of Subjective Bias in Emotion Labels}
Subjective bias exists in emotion labels, which can result in poor controlling performance for end-to-end methods. 
To prove the existence of the subjective bias in emotion labels, we compare the emotion accuracy of the center samples and that of the boundary samples. 
Specifically, all the samples of EMOPIA are firstly clustered by emotion labels to get four emotion clusters. Then, we calculate the attribute average in each emotion cluster to get the clustering center. We choose $50$ samples that are closest to the clustering center (i.e. center samples) and $50$ samples that are distant from the clustering center (i.e., boundary samples). We ask $10$ listeners to classify the samples into four emotion categories to get subjective accuracy and use the classification model to get objective accuracy. As shown in \autoref{tab:center_boundary}, both the subjective and objective accuracy in classifying the center samples is higher than that in classifying the boundary samples, which indicates that subjective bias exists in the dataset, especially in the labels of the boundary samples, and this subjective bias can hinder classification performance. We further validate this by performing t-SNE visualization of the samples in EMOPIA with central mapping and distance analysis of attribute vectors of samples from EMOPIA. The t-SNE visualization shown in \autoref{fig:t_sne_emopia} reveals that the samples in EMOPIA fail to be grouped separately. As shown in \autoref{fig:l1_distance_emopia}, the wider area between two curves indicates better performance in differentiating different groups, however, there is not much distance between the curves calculated from samples of EMOPIA. The above results further prove that subjective bias exists in emotion labels.
\paragraph{Subjective Bias Elimination}
To prove that \modelname can eliminate subjective bias in emotion labels, we compare the classification accuracy of the generated center samples and that of the generated boundary samples. Specifically, the center samples are generated by using the attribute values of the center samples extracted in the emotion-to-attribute mapping stage. Similarly, the generated boundary samples use attribute values of the boundary samples. 
As shown in \autoref{tab:center_boundary}, both the subjective and objective accuracy of the center samples are higher than those of the boundary samples, which indicates the effectiveness of the supervised clustering in the emotion-to-attribute mapping stage in eliminating subjective bias. We further validate this by performing t-SNE visualization of the samples generated by \modelname with central mapping and distance analysis of attribute vectors of samples generated by \modelname. As shown in \autoref{fig:t_sne_emogen}, the samples generated by \modelname can be clustered into four distinct groups. As shown in \autoref{fig:l1_distance_emogen}, the intra-class distance of \modelname is smaller than that of EMOPIA, which indicates that samples generated by \modelname are more similar in emotion expression in each emotion category. Besides, the area between the two curves is much wider than that of EMOPIA, which indicates better performance in differentiating samples from different emotion categories. The above results both show that \modelname can eliminate subjective bias from emotion labels.

\begin{table}[t]
    \centering
    \caption{Subjective and objective accuracy of using the center and boundary samples.}
    \begin{tabular}{c|c|c}
        \toprule
                &  Subjective  & Objective\\
         Method &  Accuracy & Accuracy\\
        \midrule
        EMOPIA (center)         &  $\bm{0.722}$ & \bm{$0.850$} \\
        EMOPIA (boundary)            &    $0.528$ & $0.655$ \\
        \midrule
         \modelname (center) & $\bm{0.675}$ &  \bm{$0.658$}  \\
         \modelname (boundary)       &  $0.300$   & $0.251$ \\
        \bottomrule
    \end{tabular}
    \label{tab:center_boundary}
\end{table}

\subsection{Comprehensive Analysis}
In this subsection, we conduct analysis experiments on different modules: 1) Emotion-to-attribute mapping methods; 2) Attribute design methods; 3) Top-$k$ attributes. More experiment implementation details can refer to Appendix \autoref{app:analysis}.

\paragraph{Emotion-to-Attribute Mapping Methods} \label{sec:analysis_first_stage}
In the emotion-to-attribute mapping stage, we need to determine a set of attribute values as the mapped attribute values for each emotion category, for which we consider the following methods: 
1) Closest: Directly using the attribute values of the sample whose attribute values are closest to the average attribute values of all the samples of the emotion. It is the default method of \modelname.
2) Center: Directly using the average attribute values of all the samples of the emotion as the mapped attribute values; 
3) K-Means: Clustering the samples of each emotion with the K-Means clustering algorithm \cite{lloyd1982least} and selecting the attribute values of the center of the largest cluster as the mapped ones. 

From the evaluation results shown in \autoref{tab:mapping}, we can see that Closest achieves better subjective and objective accuracy than Center and K-Means. Since Closest obtains attribute values out of a real sample in the dataset, it can maintain the original attribute distribution, and thus can achieve higher accuracy. On the contrary, the attribute values obtained by Center and K-Means are not from a real sample, so the value distribution deviates from a real one, which may result in poor control accuracy. As for the music quality, although the humanness score of Center and K-Means is higher than Closest, the difference is not significant. Therefore, we choose Closest as the default mapping method in the emotion-to-attribute mapping stage.

\begin{table}[t]
    \centering
    \caption{Evaluation results of different emotion-to-attribute mapping methods.}
    \begin{tabular}{c|cc|c}
        \toprule
                & \multicolumn{2}{c|}{Subjective} & Objective \\
         Method & Accuracy &  Humanness & Accuracy \\
        \midrule
        Closest         & \bm{$0.714$}  & $3.64$ & \bm{$0.658$} \\
        Center          & $0.464$       & $3.66$ & $0.657$ \\
        K-Means         & $0.607$       & \bm{$3.71$}  & $0.565$ \\
        \bottomrule
    \end{tabular}
    \label{tab:mapping}
\end{table}

\paragraph{Attributes Design}
We compare altogether four alternatives of the attribute design module: 1) Top-$100$: Using the top-$100$ attributes according to feature importance, which is the default method of \modelname; 2) Random: Selecting \num{100} attributes randomly according to attribute groups described in \autoref{sec:attr_selection}. The details of how to select these attributes are described in Appendix \autoref{app:attributes_design}. 3) Manual: Following previous work \cite{zheng2021emotionbox,tan2020music,mckay2018jsymbolic,kim2010music}, we use $17$ manually designed music attributes that are related to emotions.

\begin{table}[t]
    \centering
    \caption{Evaluation results of different attribute design methods.}
    \begin{tabular}{c|cc|c}
        \toprule
                    & \multicolumn{2}{c|}{Subjective} & Objective \\
        Method      &  Accuracy & Humanness & Accuracy \\
        \midrule
        Top-100     & \bm{$0.785$} &    \bm{$3.73$} &      \bm{$0.658$}      \\
        Random      &  $0.428$ &    $3.45$ &     $0.493$       \\
        Manual      & $0.428$ &    $3.52$ &      $0.359$       \\
        \bottomrule
    \end{tabular}
    \label{tab:attribute_selection}
\end{table}

The results are shown in \autoref{tab:attribute_selection}. We can observe that: 
1) \modelname (Top-100) improves subjective accuracy and objective accuracy by more than 35.7\% and 16.5\% separately compared with Random and Manual, which shows better controllability for our attribute design methods. The humanness score of \modelname is also superior to Random and Manual, which demonstrates the better performance of \modelname to generate high-quality music.
2) The subjective and objective emotion accuracy of Random is lower than Top-$100$, which indicates that the randomly selected attributes may not be enough to convey emotion-related information. This is reasonable since without the design to build the relationship between attributes and emotions, it is hard for the model to be trained with the controlling of emotional information.
3) The subjective and objective accuracy of Manual is lower than Top-$100$, which indicates that designing attributes through prior knowledge cannot well model the relationship between emotions and music. Therefore, we choose the top-$100$ attributes as the designed attributes of \modelname.

\paragraph{Different Top-$k$ Attributes} We further analyze the influence of the number of designed attributes (i.e., $k$) on the model performance.
Specifically, we vary $k$ in $(10,50,100,300,500)$ and evaluate the model with respect to both controllability and music quality.  

The evaluation results are shown in \autoref{tab:top_k}. We can observe that: 1) Top-$100$ achieves the highest subjective accuracy. With $k$ increasing from $10$ to $500$, the subjective accuracy increases first and then decreases, which indicates that more attributes can help improve control accuracy, yet too many can harm the controllability and this may be because they have introduced more noises; 2) The objective accuracy of top-$300$ and top-$500$ is higher than top-$100$. The reason may be that a large number of attributes can cause the mapping relationship to overfit the labeled datasets and let the model generate music very similar to the ground truth, to which the emotion classifier tends to give a more correct prediction. Due to this matter, we believe that subjective accuracy is more credible than objective one. 

3) As $k$ increases, the humanness score generally decreases, which indicates that more attributes would cause lower music quality. This is reasonable because if there are much more attributes, it would be more difficult and more data-scarce for the generative model to learn the mapping from the attributes to the corresponding music, and accordingly cause the loss of music quality. However, top-$100$ can still have relatively good quality. 
Combing the performances on both subjective accuracy and music quality, we set $k=100$ as the default value of \modelname.

\begin{table}[t]
    \centering
    \caption{Evaluation results of different top-$k$ attributes.}
    \begin{tabular}{c|cc|c}
        \toprule
                        & \multicolumn{2}{c|}{Subjective} & Objective \\
         Top-$k$ value &  Accuracy & Humanness & Accuracy \\
        \midrule
              $10$              &  $0.321$       & \bm{$3.79$}    & $0.450$    \\
              $50$              &  $0.464$       & $3.77$         &  $0.606$   \\
              $100$             &  \bm{$0.750$}  & $3.73$         &   $0.658$  \\
              $300$             &  $0.714$       & $3.46$         &   \bm{$0.790$}  \\
              $500$             &  $0.643$       & $3.49$         &    $0.784$ \\
        \bottomrule
    \end{tabular}
    \label{tab:top_k}
\end{table}

\subsection{Application on Multi-Instrument Datasets}
\label{sec:ext_topmagd}
To evaluate \modelname's ability to generate emotional music on the arbitrary dataset, we conduct experiments of \modelname on TopMAGD \cite{ferraro2018large}, which is a multi-instrument dataset containing $22535$ samples with no emotion annotations. Specifically, since the mapped attributes in the emotion-to-attribute mapping stage are determined, we only need to train the attribute-to-music generation model on TopMAGD. We use subjective accuracy, humanness, and overall as subjective metrics. For details of the subjective experiment, please refer to Appendix \autoref{app:topmagd}. We compare \modelname with the results in \autoref{tab:main_results}. 

The subjective accuracy of \modelname on TopMAGD is $0.433$, and the humanness and overall score are $3.72\pm0.17$ and $3.67\pm0.15$, respectively. We can observe that: Compared with the results in \autoref{tab:main_results}, \modelname training on TopMAGD performs better than CS and PUCT in \texttt{S1} in control accuracy and music quality. In conclusion, \modelname is able to generate music with desired emotion on multi-instrumental datasets. Generated samples are available via this link$\footnote{\url{https://emo-gen.github.io/}}$.

\section{Conclusion} \label{sec:conclusion}
In this paper, we propose \modelname, an emotional music generation system that leverages a set of emotion-related music attributes as the bridge between emotion and music. \modelname divides emotional music generation into two stages: in music-to-attribute mapping stage, \modelname map the emotion label to attribute values that can represent the general emotion by supervised clustering; in attribute-to-music generation stage, \modelname train the generative model via self-supervised learning without emotion labels. Benefiting from two stages, \modelname eliminates the subjective bias in emotion labels, so as to achieve better control accuracy. Experiment results show that \modelname is able to generate music with better emotion control accuracy and music quality compared to the previous methods. 

In the future, we will consider improving or extending \modelname in the following aspects: First, \modelname selects the sample closest to the attribute center in the emotion-to-attribute mapping stage, which may ignore the diversity of emotions. It is worth exploring how to cluster attribute vectors in fine-grained emotion classes to get more diverse emotional mapping. Second, \modelname controls the music generation with song-level attributes globally, we will further explore how to control this process dynamically to achieve emotion transitions between bar, phrase, and section levels. Finally, we expect to extend \modelname to more tasks and domains, such as emotion/style-controlled text generation. 
\bibliography{example_paper}
\bibliographystyle{icml2023}

\newpage
\appendix
\onecolumn

\section{Selected Attributes List} \label{app:attributes}
We extract $1495$ music attributes from jSymbolic. The definition of these music attributes can be found at \url{https://jmir.sourceforge.net/manuals/jSymbolic_manual/home.html}. We first train a Random Forest classifier on EMOPIA, then select $100$ attributes according to their feature importance. The first $10$ attributes are shown in \autoref{tab:first-10}.
\begin{table} [ht] 
\centering
\caption{The first ten selected attributes.} \label{tab:first-10}
\begin{tabular}{|l|} 
\hline
Note Density per Quarter Note             \\ \hline
Note Density per Quarter Note Variability \\ \hline
Total Number of Notes                     \\ \hline
Relative Note Density of Highest Line     \\ \hline
Prevalence of Long Rhythmic Values        \\ \hline
Prevalence of Very Long Rhythmic Values   \\ \hline
Average Note to Note Change in Dynamics   \\ \hline
Pitch Class Histogram\_8                  \\ \hline
Rhythmic Value Histogram\_10              \\ \hline
Vertical Interval Histogram\_43           \\ \hline
\end{tabular}
\end{table}

For more details about selected attributes, please refer to \url{https://emo-gen.github.io/}.
\section{Details of Experiments}
\subsection{Comparison with Previous Methods} \label{app:main_results}
We invite $15$ participants to evaluate $32$ songs which consist of $4$ emotion categories for each setting and method. The participant needs to rate music samples on a five-point scale with respect to 1) Valence: Is the music piece negative or positive; 2) Arousal: Is the music piece low or high in arousal; 3) Humanness: How similar it sounds to a piece composed by a human; 4) Overall: An overall score.  For objective metrics, we apply the emotion classifier to classify the generated $1000$ samples for each method, then we calculate objective accuracy by comparing the emotion input for generating music with the predicted emotion class by this classifier.

\begin{table*}[ht]
\centering
\caption{Detail results of subjective evaluation. Metrics hv, lv, ha, la stand for high valence, low valence, high arousal, and low arousal respectively. For all metrics, we report mean opinion scores and $95$\% confidence interval.}
\resizebox{\linewidth}{!}{
\begin{tabular}{ccccccc}
\toprule
  Setting & Method                     & hv$\uparrow$  & lv$\downarrow$ & ha$\uparrow$ & la$\downarrow$ \\ \midrule
EMOPIA     & Ground truth                        &  $3.83\pm 0.40$   & $2.83\pm0.53$   & $4.13\pm0.34$   & $2.53\pm 0.42$  \\ \midrule
\multirow{3}{*}{\thead{\texttt{S1}: Pre-training on Pop1k7+LMD-Piano, \\fine-tuning on EMOPIA}} & CS                          &  $3.00\pm0.44$   & $3.23\pm0.40$   & $3.9\pm0.31$   & $2.93\pm0.43$   \\
                   & PUCT                           &   $3.20\pm0.35$  & $3.20\pm0.39$   & $3.40\pm0.40$   & $3.07\pm0.41$   \\
                   & \modelname                     &  \bm{$3.43\pm0.44$}   & \bm{$2.40\pm0.44$}   & \bm{$4.27\pm0.19$}   & \bm{$1.77\pm0.31$}   \\ \midrule
\thead{\texttt{S2}: Training on Pop1k7+LMD-Piano+EMOPIA}               & \modelname                     &  $3.30\pm0.47$   & $2.57\pm0.48$   & $4.17\pm0.29$   & $1.90\pm0.35$  \\ 

\bottomrule
\end{tabular}}
    \centering
    \label{apx:tab:main:sub}
\end{table*}

To further compare the results of each method, we calculated the average valence and arousal scores, respectively. Detail evaluation results for valence and arousal are shown in \autoref{apx:tab:main:sub}. As we can see: 
\begin{enumerate}
    \item In \texttt{S1}, \modelname outperforms CS and PUCT in hv,lv,ha and la. Thus, \modelname controls emotion better than CS and PUCT under this setting.  
    \item \modelname in Setting 2 controls emotion better than CS and PUCT in Setting 1. Therefore, benefiting from the two-stage framework, \modelname can achieve good performance in arbitrary datasets with no annotations.
\end{enumerate}

\subsection{Verification on Eliminating Subjective Bias} \label{app:bias}
Considering that each emotion in EMOPIA contains approximately $250$ samples, we select $50$ samples from the center and boundary respectively for each emotion category. And for \modelname, we generate $1000$ samples for the center and boundary separately. Each emotion category contains $250$ samples. We invite $10$ participants to classify the music samples into $4$ emotional quadrants according to Russell's 4Q model\cite{russell1980circumplex}. Each participant evaluates $16$ samples randomly sampled from 1) Center and boundary of EMOPIA; 2) Center and boundary generated by \modelname. Samples are evenly distributed in four emotion categories. For objective metrics, we apply the emotion classifier to predict the emotion label of  samples from EMOPIA and \modelname, then the objective accuracy can be obtained.

\subsection{Comprehensive Analysis}\label{app:analysis}
For each compared module and method, we generate $1000$ samples with $250$ for each emotion category. Then we invite $7$ participants to 1) Classify the sample into four emotional categories based on Russell's 4Q model; 2) Rate the humanness score of the sample on a five-point scale. The higher the score is, the more realistic the sample is to a human-composed one. Each participant receives $44$ samples evenly distributed in $4$ emotion categories, which are divided into $3$ groups: 1) Group for $3$ compared methods in emotion-to-attribute mapping, which consists of $12$ samples; 2) Group for $3$ compared methods in attribute design, which consists of $12$ samples; 3) Group for $5$ values of top-$k$, which consists of $20$ samples. For objective metrics, we apply the emotion classifier to predict the emotion labels of $1000$ generated samples and objective accuracy is calculated similarly to Appendix \autoref{app:main_results}.  
\paragraph{Details of Attribute Design} \label{app:attributes_design}
We present details about another two attribute design methods in \autoref{tab:attribute_selection}:
\begin{enumerate}
    \item Random: Since jSymbolic divides attributes into seven groups (please refer to \url{https://jmir.sourceforge.net/manuals/jSymbolic_manual/home.html} for detail), we select $100$ attributes on average according to the $7$ groups.
    \item Manual: Following previous work \cite{zheng2021emotionbox, tan2020music, mckay2018jsymbolic,kim2010music}, we choose pitch class histogram, note density, rhythm density, and mean pitch/duration/velocity as manually designed attributes related to music emotion, which form a $17$-dimensional attribute vector.
\end{enumerate}

\subsection{Application on Multi-Instrument Datasets} \label{app:topmagd}
We keep the emotion-to-attribute mapping stage unchanged and train the attribute-to-music generation stage on TopMAGD. We apply \modelname to generate $1000$ samples with $250$ for each emotion category. And we invite $15$ participants to rate each sample by Valence, Arousal, Humanness, and Overall score similar to \autoref{app:main_results}. Each participant evaluates $4$ samples which consist of $4$ emotion categories. Then we report subjective accuracy (the same as the calculation rule in \autoref{app:main_results}), humanness, and overall score as subjective evaluation metrics. 

\subsection{Discussion on Output Diversity}
\modelname uses only one set of attributes to represent each emotion, which will limit the diversity.
The emotion of music contains two levels, one is the emotion felt by most people (i.e., general emotions), and the other is the emotion felt by individuals (i.e., personalized emotions). General emotions are not as diverse as personalized emotions. In this paper, we mainly consider controlling the music generation with general emotions not influenced by subjective bias. 
However, if needed, EmoGen can also achieve more emotion diversity by mapping other emotions into more sets of attribute values in the emotion-to-attribute mapping stage.


\end{document}